# Cooling in a Bistable Optical Cavity

Mark Y. Vilensky, Yehiam Prior, and Ilya Sh. Averbukh

Department of Chemical Physics
Weizmann Institute of Science
Rehovot, Israel 76100

We propose a generic approach to nonresonant laser cooling of atoms/molecules in a bistable optical cavity. The method exemplifies a photonic version of Sisyphus cooling, in which the matter-dressed cavity extracts energy from the particles and discharges it to the external field as a result of sudden transitions between two stable states.

Laser cooling is an extremely successful method for creating the ultra-cold ensembles of atoms[1-3]. Doppler cooling (a crucial component of laser cooling for all but lowest temperatures) relies on the multiple absorption-emission cycles in a closed system of levels. It is limited to alkali metals and to a small number of other atomic species. This makes laser cooling of molecules very difficult[4], although there are a few proposals and experimental demonstrations of cooling molecules[5]. Moreover, the effects of radiation trapping and excited state collisions limit the final achievable phase-space density in cooling methods based on resonant light absorption.

Non-resonant laser cooling methods such as cavity cooling[6,7] and stochastic cooling[8-11] were shown to be potentially useful for cooling dense samples of the trapped particles down to the mK region. Cavity cooling is based on the generation of viscous-type friction force imposed on the atoms as they move inside a leaky cavity. Recently, single atom cavity cooling was experimentally demonstrated[12,13], and further improvements by introducing of a linear feedback were suggested[14]. The related subject of radiation pressure cooling of microscopic mechanical resonators, such as micromirrors[15,16] and microlevers[17] has attracted significant attention recently, potentially leading to the observation of the quantum ground state of a micro-mechanical system.

In this paper we propose a generic scheme for non-resonant atomic/molecular cooling in *bistable* optical cavities. The cooling mechanism is of Sisyphus type[18], in which the cavity mode (not the atom) performs sudden transitions between two stable states. Conventional cavity cooling relies on non-adiabatic effects in the cavity-atom interaction, and requires high finesse cavities. In contrast, our technique best operates in the opposite limit of "bad cavity" and slowly moving particles, so that the cooling rate does not experience deterioration at low atomic velocities.

Consider an optical resonator supporting a standing wave $\cos(2\pi x/\Lambda)\exp(-i\omega_R t)$ (where $\Lambda$ is the spatial period) and $\omega_R$ is the bare cavity resonance frequency. The resonator is externally excited by an incident plane wave $E_i \exp\left[-i\omega(t - x/c)\right] + c.c$, which is nearly resonant with the cavity mode $\Delta_C \ll \omega_R$ (where $\Delta_C \equiv \omega_R - \omega$). A point-like polarizable particle moving inside the resonator contributes an effective refractive index depending on the mode function at the particle's instantaneous position. As a result, the cavity resonant frequency experiences a position dependent shift, $U(x) = U_0 \cos^2(2\pi x/\Lambda)$, where $U_0 = \omega \operatorname{Re}(\alpha)/(\varepsilon_0 V)$, $\alpha$ is polarizability of the particle, $V$ is the mode volume, and $\varepsilon_0$ is the permittivity of free space (see e.g. [19]). Thus, the intra-cavity field becomes strongly coupled to the particle motion. The complementary aspect of this coupling is that the dipole force felt by the particle depends on the local intensity of the field inside the resonator. This generic interplay between the cavity field dynamics and the particle motion forms the basis for all cavity cooling schemes. The combined system dynamics is given by the following set of coupled equations for the cavity field amplitude and the particle velocity and position:

$$\frac{1}{\kappa}\frac{dE}{dt} = \frac{2E_i}{\sqrt{T}} - E\left[1 + \frac{\gamma(x)}{\kappa} + i\frac{\Delta_C}{\kappa} - i\frac{U(x)}{\kappa}\right]$$

$$\frac{dv}{dt} = -\frac{2\pi \operatorname{Re}(\alpha)}{m\Lambda}|E|^2 \sin\left(\frac{4\pi x}{\Lambda}\right) \quad (1)$$

$$\frac{dx}{dt} = v$$

Here $E \exp(-i\omega t) \cos(2\pi x/\Lambda) + c.c$ is the field in the cavity. The decay rate of the cavity field is denoted by $\kappa = Tc/2L$, where $T$ is the transmission coefficient of the cavity mirrors and $L$ is the cavity length. In addition, $\gamma(x) = \omega \operatorname{Im}(\alpha)/(\varepsilon_0 V) \cos^2(2\pi x)$ is the scattering rate of the atom, and $m$ is the mass of the particle.

In the adiabatic limit, when the particle's velocity is small enough ($v \ll \kappa\Lambda$), the cavity field responds to the motion and attains steady state for any position of the particle. The steady-state intensity of the cavity field can be extracted from the first equation of the system of Eqs. (1):

$$|E|^2 = \frac{4|E_i|^2}{T\left[(1+\gamma(x)/\kappa)^2 + \Delta^2(x)/\kappa^2\right]}, \quad (2)$$

where $\Delta(x) = \Delta_C - U(x)$ is the effective detuning of the cavity which depends on the instantaneous position of the particle. Substitution of Eq. (2) into the second of Eqs. (1) reduces the problem to that of the motion of a particle in an effective periodic potential. Because of the conservative character of this motion, energy that is lost while climbing a potential hill is regained while sliding down on the other side. Therefore, no net energy loss is possible in the adiabatic limit, and the cooling process requires non-adiabatic effects. All previously discussed cavity cooling schemes[6, 7] rely on high finesse cavities that ensure a retarded response of the internal field to the changing position of the particle. There is however, a notable example for a profound non-adiabatic effect in "bad" (low finesse) cavities, which is related to the phenomenon of optical bi- or multi-stability[20]. Under certain conditions (see below), a driven optical resonator has two co-existing stable steady states. A sharp transition between these states may happen even at infinitely slow variation of the resonator parameters, leading to a rapid passage from high to low cavity field (or vice versa). In what follows, we demonstrate a novel scheme for cooling using a bistable cavity.

We now consider the situation when the pump frequency is far detuned from any atomic transition, so that the scattering loss $\gamma(x)$ may be neglected. We introduce an external feedback circuit (discussed below) such that the external pump field intensity $I_i = |E_i|^2$ depends on the field intensity $I = |E|^2$ inside the cavity. In this case, expression (2) becomes an equation from which the steady-state intra-cavity intensity should be found:

$$I\left[1 + \Delta^2(x)/\kappa^2\right] = \frac{4I_i(I)}{T} \qquad (3)$$

For a given dependence $I_i(I)$, equation (3) may be solved for $I$, e.g. graphically (see Fig. 1a) by considering the intersection of a dashed line (LHS in Eq. (3)) and the feedback curve (RHS in Eq. (3)). We assume the latter to have a sigmoid shape (solid line in Fig. 1a). The slope of the straight line depends on the particle position and varies from $1 + (\Delta_C - U_0)^2/\kappa^2$ to $1 + \Delta_C^2/\kappa^2$ (and back) as the particle moves. Depending on the parameter values, the feedback curve and the straight line may have one or three intersections (each corresponding to a solution of Eq.(3)). In the case of triple intersection, linear stability analysis shows that only two of the solutions (shown by solid circles in Fig. 1a) are stable[20].

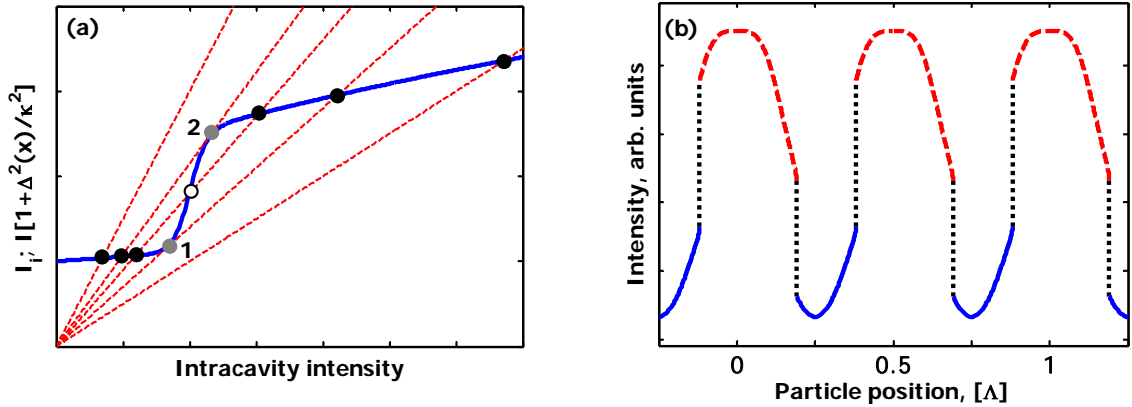

**Fig. 1** (color online). (a) Solutions of Eq. (3) corresponding to different particle positions. Stable solutions are shown by solid circles, unstable one is shown by an empty circle. Grey circles (numbered 1 and 2) show the critical points at which the solution jumps from one stable brunch to another. (b) Intensity of the intra-cavity field versus particle position. The solid (blue online) curve shows the solution corresponding to the lower branch of the feedback curve in figure (a), the dashed (red online) curve corresponds to the upper branch. Dotted lines show the intensity switching.

Let us follow the motion of a high field seeking ($U_0 > 0$) particle over one period of the standing wave. We consider the case of $\Delta_C > 0$ (pump frequency is lower than $\omega_R$), and relatively weak coupling, $U_0 < \Delta_C$. When the particle is at an antinode of the standing wave (e.g., $x = 0$), the cavity detuning $\Delta(x)$ is minimal, and the amplitude of the internal field is high (see Fig. 1b). As the particle moves toward the node of the standing wave mode, the slope of the straight line in Fig. 1a gradually increases from $1 + (\Delta_C - U_0)^2/\kappa^2$ to the maximal value of $1 + \Delta_C^2/\kappa^2$. When the particle is close to the antinode, the steady-state solution follows the upper branch of the sigmoid curve. At a certain particle position, the straight line in Fig. 1a detaches from the sigmoid feedback curve, and the intra-cavity intensity drops down to the solution at the lower branch of this curve. After the particle reaches the node, the slope starts decreasing back from its maximal value $1+\Delta_C^2/\kappa^2$. The steady state solution follows the lower branch of the feedback curve until the multiple intersections disappear, and the intra-cavity intensity jumps to its high value. As follows from Fig. 1a, the field switches up and down at non-equivalent particle positions (hysteresis effect that is well known for bistable resonators[20]). Such a pair of sudden jumps occurs twice on every period of the standing wave mode. When the particle moves adiabatically over the standing wave, it gains kinetic energy when sliding down the effective potential and loses it when climbing up. However, in the present case, the loss and gain are not equal because of the hysteresis effect. Depending on the relative sign between the particle polarizability (red or blue detuning from the atomic resonance) and cavity detuning $\Delta_C \equiv \omega_R - \omega$, the particle can (on average) either lose or gain energy. For the parameters chosen above, ($\Delta_C > 0, U_0 > 0$, and $U_0 < \Delta_C$) the particle slows down. Moreover, the particle loses the same amount of energy on passing every single spatial period of the standing wave until it is trapped by the optical potential. The average stopping force acting on the particle is, therefore, constant and does not depend on the velocity. It is noteworthy that while in other cavity cooling schemes[6, 7, 19] the resonator introduces viscous-type friction (proportional to the particle velocity), our bistable resonator causes a "dry friction" force that is velocity independent. Hence, this scheme promises a more efficient slowing the particle.

To estimate this force, we consider a simple model in which the sigmoid feedback curve is replaced by a step-like function: $|E_i|^2 = I_1$ for $|E|^2 < I_{sw}$ and $|E_i|^2 = I_2$ for $|E|^2 \geq I_{sw}$, where $I_{sw}$ is the cavity intensity value at which the intensity of the incident field switches from $I_1$ to $I_2$. The steady-state intra-cavity intensity is then double-valued: $|E|_n^2(x) = 4\kappa^2 I_n/[T(\kappa^2 + \Delta^2(x))]$, $(n=1,2)$. Under these conditions, the effective potential energy felt by the particle has two branches as well:

$$W_n(x) = \int^x dx' \frac{\varepsilon_0 V}{\omega} \frac{dU(x')}{dx'} |E|_n^2(x')$$

$$= \frac{4V\varepsilon_0 I_n \kappa}{\omega T} \arctan(\Delta(x)/\kappa), \quad (n=1,2). \tag{4}$$

For a slowly moving particle, the energy loss per half a period of the mode function is given by,

$$\Delta E = -\frac{4V\varepsilon_0 \kappa (I_2 - I_1)}{\omega T}\left[\arctan\left(\frac{\Delta_2}{\kappa}\right) - \arctan\left(\frac{\Delta_1}{\kappa}\right)\right], \tag{5}$$

where $\Delta_1$ and $\Delta_2$ are the critical detuning values at which the straight lines touch the feedback curve: $(\Delta_{1,2}/\kappa)^2 = (4I_{1,2} - TI_{sw})/(TI_{sw})$. The average stopping force is given by $F_{stop} = 2\Delta E/\Lambda$. For $U_0/\kappa \ll 1$, the particle motion best modulates the cavity transmission at $\Delta_C \sim \kappa$, i.e. at the slope of the resonant Lorentzian curve, which also holds for the regular cavity cooling[21]. By adjusting the feedback curve such that the intra-cavity intensity switches up when the particle is in the field antinode, and the intensity goes down when the particle approaches the node, the average stopping force in the limit of $U_0/\kappa \ll 1$, and $\Delta_C = \kappa$, is:

$$F_{stop} = -\frac{4V\varepsilon_0 \kappa}{\Lambda \omega T}\left(\frac{U_0}{\kappa}\right)^2 I_0, \tag{6}$$

where $I_0 = (I_1 + I_2)/2 \approx I_1 \approx I_2$ is the average input intensity.

Figures 2a and 2b present two examples of the bistable cavity cooling for weak ($U_0/\kappa = 0.1$) and strong ($U_0/\kappa = 1.33$) atom-cavity coupling, respectively. Both figures depict the simulated temporal evolution of the velocity of a single particle moving in a bistable cavity (lower curve) and compare it to cavity cooling without the feedback (upper curve). In these simulations, the sigmoid feedback step function was represented by a smooth continuous function:

$$I_i = I_0 + (\Delta I/2)\tanh\left[a(I - I_{sw})T/I_0\right], \qquad (7)$$
$$I_0 = (I_1 + I_2)/2, \; \Delta I = I_2 - I_1$$

The parameter $a$ controls the steepness of the feedback curve step. Cavity cooling without feedback corresponds to $\Delta I = 0$.

The dimensionless parameters for Fig. 2a were chosen as $U_0/\kappa = 0.1$, $\Delta I/I_0 = 0.13$, $TI_{sw}/I_0 = 0.53$, $a = 50$ and $\mathrm{Re}(\alpha)I_0/(\kappa^2 \Lambda^2 Tm) = 2.5 \cdot 10^{-5}$. The latter parameter is actually proportional to the ratio of the polarization energy and a typical kinetic energy of a particle moving with the velocity $v = \kappa\Lambda$. The chosen parameters correspond to a particle having mass of Rb atom and pumped far off-resonance (about 70 natural linewidth below the atomic transition). The initial velocity of the particle is $\sim 2500$ recoil velocities, $v_{rec}$ of the rubidium atom. The cavity is 500 $\mu$m long with the mode diameter of 15 $\mu$m and $\Delta_C = \kappa = 1.2 \cdot 10^8 \, \mathrm{s}^{-1}$, pumped by 5 nW light of 1 μm wavelength. At the initial stage, the non-adiabatic effects are prevailing and the velocity of the particle decreases in a way similar to the conventional cavity cooling[6, 7, 19]. As the particle slows down, the bistability "kicks in", and the velocity starts to decrease linearly with time (see Fig. 2a). The numerical value of the decelerating force is in good agreement with the analytical estimate, Eq. (6). The insert in Fig. 2a shows the time dependence of the cavity field in the domain of constant deceleration. Jump-like changes in the field intensity are clearly seen as discussed above (compare with Fig. 1b). The deceleration stops as the particle reaches the velocity of $\sim 250 \, v_{rec}$ and becomes trapped by the optical potential. For comparison, the upper curve in Fig. 2a demonstrates the same process in a cavity without the feedback and is evidently less efficient under the same conditions.

The parameters for the Fig. 2b were chosen to resemble qualitatively the conditions of the conventional cavity cooling experiment[12], i.e. strong atom-cavity coupling, exact cavity resonance ($\Delta_C = 0$), and blue detuning from atomic transition (negative polarizability). The following set of parameters was used: $U_0/\kappa = -1.33$, $\Delta I/I_0 = 0.95$, $TI_{sw}/I_0 = 0.53$, $a = 10$, and $\mathrm{Re}(\alpha)I_0/(\kappa^2 \Lambda^2 Tm) = -5 \cdot 10^{-5}$. An efficient deceleration in the bistable regime is seen, which is faster than that of Fig. 2a due to stronger coupling.

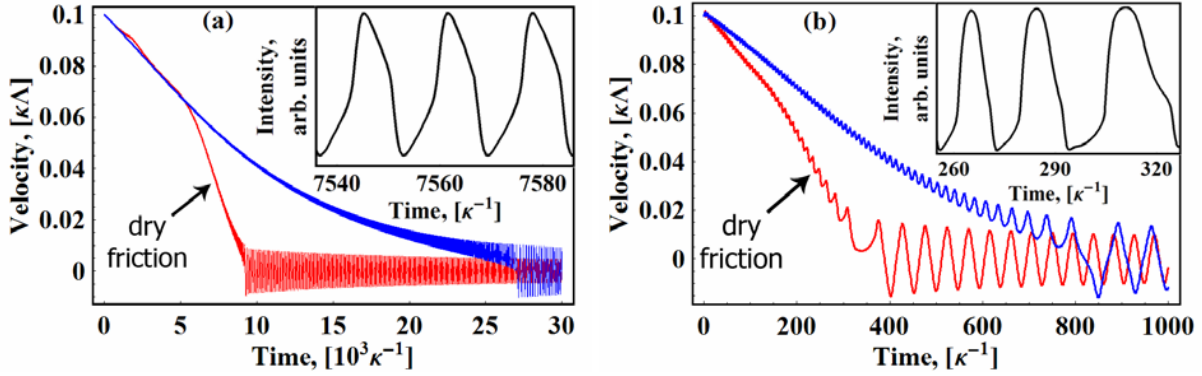

**Fig. 2** (color online). Particle deceleration in a bi-stable cavity – lower curve (red online) in the (a) weak, and (b) strong atom-cavity coupling regimes. For comparison, the upper curve (blue online) shows particle slowing in the absence of feedback $(\Delta I = 0)$. The inserts show the time dependence of the cavity field intensity in the domain of constant deceleration.

The finite response time of the feedback loop determines the upper limit of the initial velocity. In particular, a finite measurement time, $t_m$ is needed to distinguish the particle-induced variations in the cavity output ($\Delta I$ in the above model, Eq.(7)) from the Poissonian shot noise in the mean transmitted signal (proportional to $I_0$). The required condition is $t_m > (I_0/\Delta I)^2 \hbar\omega/P_0$ (where $P_0$ is mean incident light power). The particle displacement during the measurement has to be smaller than the period of the cavity mode, $vt_m/\Lambda \ll 1$. For the examples presented in Figures 2a and 2b this ratio is $0.1$ and $0.02$, respectively. Assuming also that time duration of 1 ns is required for an electro-optic circuit to switch between the two levels of the feedback step; the velocity should be less than 50 m/s in order to neglect the particle displacement in a 0.5 μm period of the optical potential.

The next step was to analyze cooling of an ensemble of $N$ particles coupled to a single mode of a bistable resonator, under conditions of blue atomic detuning $\Delta_C = 0$, $U_0/\kappa = -0.7$, and $\text{Re}(\alpha)I_0/(\kappa^2\Lambda^2 Tm) = -2.5\cdot 10^{-6}$. The feedback curve was chosen in the form of a smoothen step (as in Eq. (7)) with the parameters adjusted to ensure frequent switching between two stable states, which results from stochastic modulation of the cavity detuning due to the particles' motion. Fig. 3a presents the results of a direct simulation of cavity cooling for $N = 5$ particles that are initially randomly dispersed inside the cavity with a Gaussian distribution in velocity with zero mean and dispersion of $0.016\kappa\Lambda$. In Fig. 3a, the evolution of the velocity variance as a function of time is depicted for both conventional (dashed) and bistable (solid) cavity cooling. A series of sharp random steps is observed in the curve describing the decrease of the average kinetic energy in the bistable regime. Each step is correlated with a "jump" in the intra-cavity field intensity. As appears from Fig. 3a, bistable cavity provides more efficient cooling compared to the cooling in the same cavity but without the feedback. Figure 3b presents similar results for $N = 25$ particles. The observed slowing down in the cooling rate with $N$ is typical to regular cavity cooling techniques[22], and to stochastic cooling methods[8-11] to which our approach belongs as well. For the conventional cavity cooling this problem has been resolved by a proper re-adjusting the parameters with the number of atoms[22], and the achieved calculated cooling rate was reported to be independent of $N$[23]. A detailed analytical and numerical study of ensemble cooling in a bistable cavity, its dependence on the external parameters and scaling with $N$ is currently under study, and will be reported elsewhere.

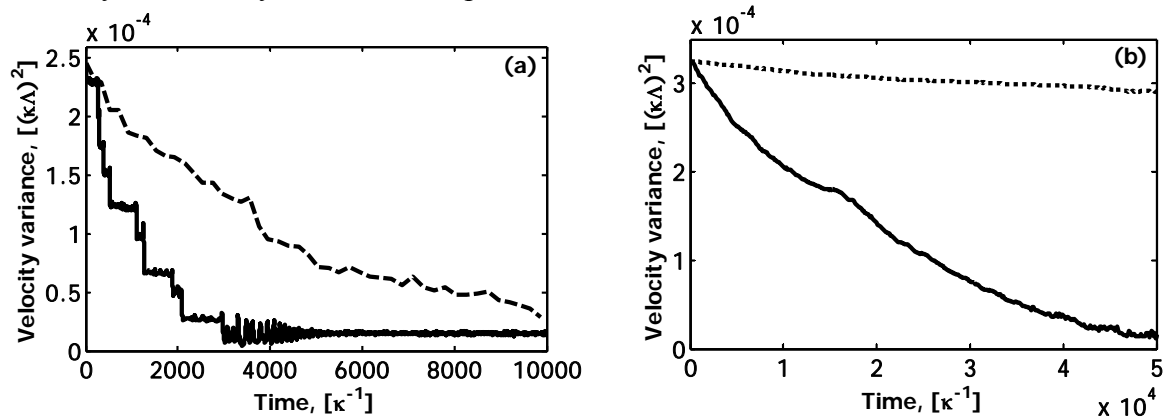

**Fig. 3.** Evolution of the velocity variance of the ensemble of particles during cooling in a bistable cavity (solid line) and in conventional cavity (dashed line). (a) $N = 5$ and (b) $N = 25$. The curves in (b) are numerically smoothened.

In conclusion, we have presented a new approach to non-resonant laser cooling of atoms and molecules based on their interaction with a bistable cavity. The cooling mechanism presents a photonic version of Sisyphus cooling, in which the conservative motion of the particles (atoms or molecules) is interrupted by sudden transitions between two stable states of the cavity mode. The mechanical energy is extracted due to the hysteretic nature of those transitions. The bistable character of the cavity may be achieved by an external feedback loop (like the one considered in the present paper) or by means of additional nonlinear intracavity optical elements (saturable absorber, nonlinear dispersion, etc. [20]). In contrast to the conventional cavity cooling[6,7], in which atoms experience a viscous-type force, bistable cavity cooling imitates "dry friction", and stops atoms much faster. Our technique operates in the "bad cavity" limit and preserves its efficiency at low particle velocities. Classically, the limit to this cooling mechanism is set by the trapping of particles in the optical potential. Quantum effects, which are not included in the present discussion, will certainly limit the cooling near the recoil limit, and are the subject of the ongoing studies. Finally, we expect that in analogy to the methodology presented here, bistable cavity cooling may also be advantageous for cooling of micro-mechanical resonators[15-17].

This work was supported in part by German-Israeli Foundation for Scientific Research and Development.